KEYSTONE

# Sea Change in Software Development:

## *Economic and Productivity Analysis of the AI-Powered Developer Lifecycle*

**Thomas Dohmke**
GitHub

**Marco Iansiti**
Harvard Business School
and Keystone.AI

**Greg Richards**
Keystone.AI

# Findings

*01.*

**Less than a year after its general availability, GitHub Copilot is turbocharging developers writing software.** Analysis on a large sample of GitHub Copilot users (n = 934,533) reveals a sizable productivity impact. On average, users accept nearly 30% of code suggestions and report increased productivity from these acceptances. Furthermore, this productivity impact increases with time, and the benefits are greatest for less experienced users.

---

*02.*

**We estimate these generative AI developer productivity benefits could boost global GDP by over $1.5 trillion USD by 2030 by helping to meet growing demand for software.** These estimates are conservative; they are moment-in-time projections that do not account for the increased demand for software development due to its greater efficiency and continued digital transformation that will arise from generative AI adoption. This analysis presents a novel perspective, additive to other projections of generative AI's expected economic impact reported by other sources.

---

*03.*

**The global landscape of players working on generative AI is diverse, including big tech, start-ups, academia, and individuals.** Open source activity on generative AI has seen an exponential increase compared to previous years, based on an analysis of GitHub repositories and commits. Findings suggest that the open source ecosystem, particularly in the United States, is driving generative AI software innovation. Individual developers are leading the majority of such repositories on GitHub.

# Generative AI redefines developer productivity

Despite its relatively recent introduction as a developer tool, generative AI is already having a dramatic impact on software development. Our own analysis examines innovation, specifically measuring generative AI's impact on developer productivity and learning with GitHub Copilot, an AI pair programmer that suggests code and entire functions in real time.[1]

Our work builds on previous research. A controlled experiment in 2022 asked developers to implement an HTTP server in JavaScript in a timed test in order to test the efficacy of GitHub Copilot. Developers were randomly assigned to a control group or a treatment group, and participants in the treatment group were given access to GitHub Copilot. The experiment revealed an important finding: Developers using GitHub Copilot implemented the server 55.8% faster than developers in the control group.[2]

A survey of users further investigated the effect of code generation on key productivity measures, including perceived productivity, satisfaction, activity, communication, and efficiency.[3,4] This revealed a strong correlation between the acceptance rate of GitHub Copilot recommendations and reported productivity and satisfaction. Developers intimately understood their work, and they reported enhanced productivity that correlated with acceptance rate.[5]

---

1 GitHub Copilot suggests source code recommendations based on context, and is powered by OpenAI's GPT-3.5 Turbo, which was trained on a diverse range of internet text, including books, articles, websites, and publicly available source code.

2 Peng, Sida, et al. (2023). "The impact of AI on developer productivity: Evidence from GitHub Copilot". arXiv preprint arXiv:2302.06590. https://arxiv.org/pdf/2302.06590.pdf.

3 Ziegler, A., Kalliamvakou, E., Li, X. A., Rice, A., Rifkin, D., Simister, S., & Aftandilian, E. (2022, June). "Productivity assessment of neural code completion". In Proceedings of the 6th ACM SIGPLAN International Symposium on Machine Programming (pp. 21–29). https://dl.acm.org/doi/pdf/10.1145/3520312.3534864.

4 Kalliamvakou, E. (2023, June). "Research: quantifying GitHub Copilot's impact on developer productivity and happiness" accessed at https://github.blog/2022-09-07-research-quantifying-github-copilots-impact-on-developer-productivity-and-happiness/ on 23rd June, 2023.

5 There is ongoing research to understand if there may be any overreliance from developers on acceptance rates.

Now that GitHub Copilot is out of technical preview and has been in the field for a year, we can study how it impacts developers over a longer period of time. The following research thus builds on earlier studies to find evidence of the large and growing benefit of GitHub Copilot. Informed by this earlier research, we are using acceptance rate as an indicator of impact and productivity.

By examining GitHub Copilot telemetry, we found that users accept an average of 30% of code suggestions, representing real productivity gains. We also discovered two additional findings:

- *The productivity benefit increases over time.*[6]
- *Less experienced developers have a greater productivity benefit.*[7]

Figure 1: **GitHub Copilot's effect increases over time**

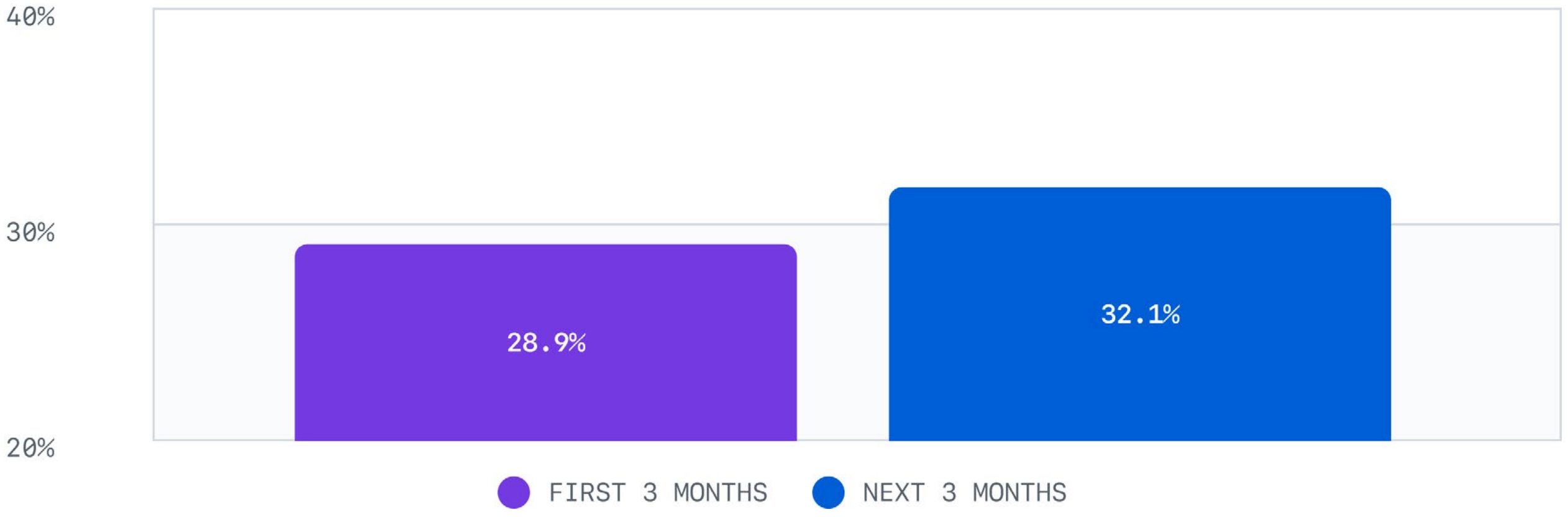

GitHub Copilot has a sizable productivity impact on developers that grows over time. In the first three months, developers accept approximately 30% of code recommendations. As developers overcome any learning curves and become more comfortable with the tool, they use it for even greater impact and accept more code suggestions, as shown in Figure 1. These results are robust to alternative econometric specifications.[8] They are also consistent

[6] This analysis was performed on all developers who used GitHub Copilot and received recommendations during the time period.

[7] This analysis was performed on developers who had repository activity on the GitHub platform prior to their use of GitHub Copilot. Less experienced developers were ones who had less average activity than the median measure of activity in the sample of developers.

[8] A regression analysis on Total Accepts is presented in the Appendix.

with and extend previous experiments and related productivity research on generative AI.[9,10]

Our analysis speaks to broader trends in developer innovation around generative AI. Developers are directly responsible for software innovation—and our analysis shows that this movement around generative AI and its impact on productivity will influence every sector of the global economy.

## The impact is large and accelerating

Figure 2: **Increase in acceptance rate of GitHub Copilot recommendations over time**

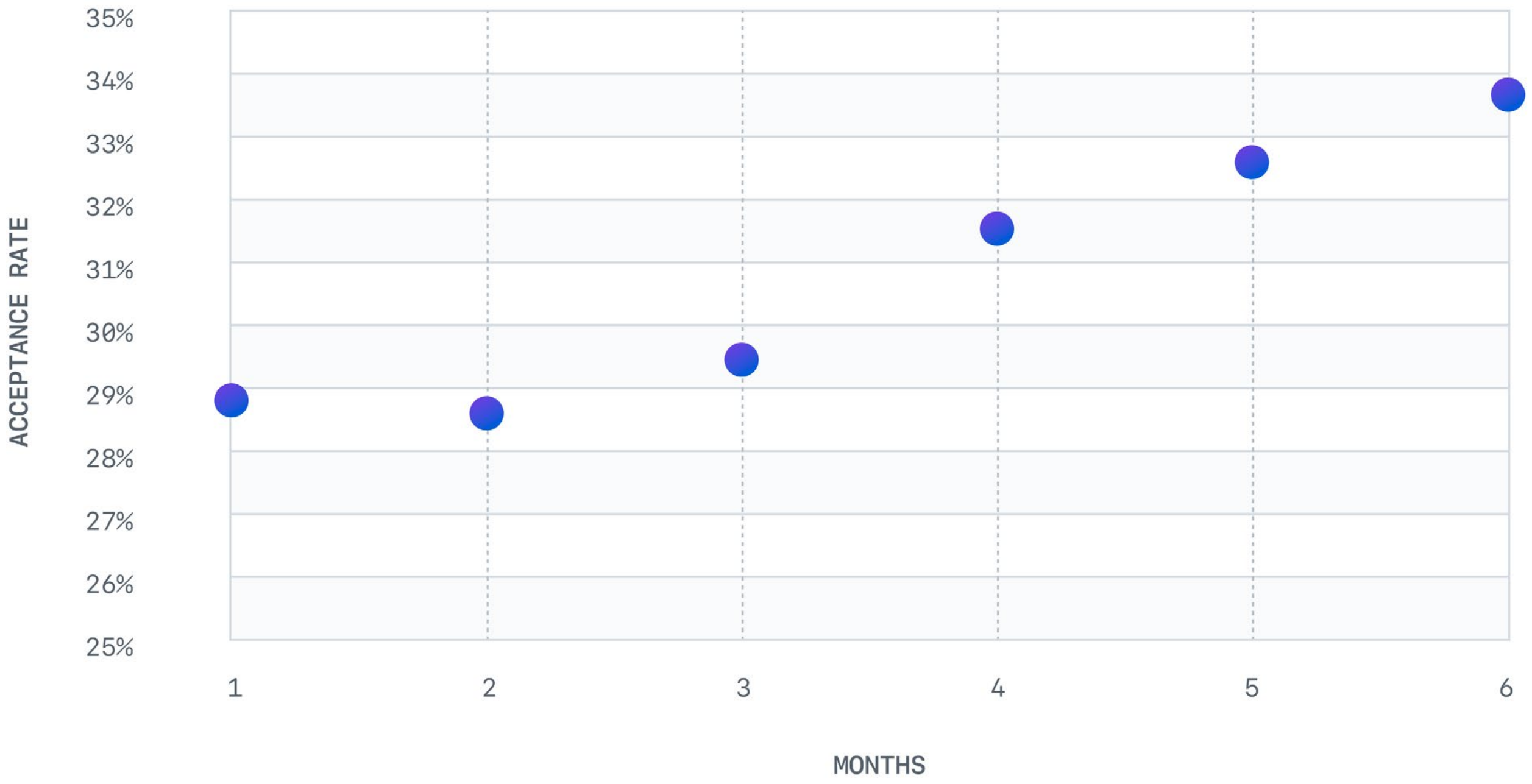

Figure 2 provides more detail on GitHub Copilot's impact over several months. Interestingly, the learning process does not appear to be diminishing after six months, which shows potential for even greater impact over time. This is important evidence for the positive impact of the tool, as developers continue to trust it more with increased usage and feedback on coding results. This is a novel, breakthrough finding. It suggests that GitHub Copilot has a large runway to continue its impact on developer productivity, as users become more accustomed to developing software with it.

[9] Peng, Sida, et al. (2023). "The impact of AI on developer productivity: Evidence from GitHub Copilot". arXiv preprint arXiv:2302.06590. https://arxiv.org/pdf/2302.06590.pdf.

[10] Noy, Shakked, and Whitney Zhang. (2023). "Experimental evidence on the productivity effects of generative artificial intelligence". SSRN 4375283. https://economics.mit.edu/sites/default/files/inline-files/Noy_Zhang_1_0.pdf.

Accepting code completion suggestions is beneficial for developers as it allows them to finish writing a code block faster and can save time searching for less-commonly used syntaxes. Acceptance rate captures code immediately adopted from GitHub Copilot suggestions based on the active choice of the developer. This is confirmed in survey research, which finds that developers' statements of productivity are closely tied to acceptance rate.[11]

Our results thus provide strong evidence that GitHub Copilot greatly impacts developer productivity across a very broad range of use cases. This is consistent with many interviews we conducted, though the precise productivity effects will vary by use case and warrants further experimental research in field settings.

## Productivity gains for less experienced developers

Next, we grouped developers into quintiles based on their experience, as measured by their average number of repository actions on the GitHub platform prior to their use of GitHub Copilot. In our study, as shown below, we found that acceptance rate is higher in this group (Figure 3).

For example, the acceptance rate for the bottom quintile is 31.9%, whereas the acceptance rate for the top quintile is 26.2%. Developers in the bottom half of repository activity have a greater acceptance rate for GitHub Copilot suggestions than those in the top half.[12] This result is consistent with earlier research on skill benefits from generative AI.[13]

---

[11] Ziegler, A., Kalliamvakou, E., Li, X. A., Rice, A., Rifkin, D., Simister, S., & Aftandilian, E. (2022, June). "Productivity assessment of neural code completion". In Proceedings of the 6th ACM SIGPLAN International Symposium on Machine Programming (pp. 21–29). https://arxiv.org/pdf/2205.06537.pdf.

[12] Regression analysis demonstrates a statistically significant relationship between less experience prior to adopting GitHub Copilot and a higher volume of accepted recommended lines of code.

[13] Brynjolfsson, Erik, Danielle Li, and Lindsey R. Raymond. (2023). "Generative AI at Work". No. w31161. National Bureau of Economic Research https://mitsloan.mit.edu/shared/ods/documents?PublicationDocumentID=9765); Peng, Sida, et al. "The impact of ai on developer productivity: Evidence from GitHub Copilot". arXiv preprint arXiv:2302.06590 (2023). https://arxiv.org/pdf/2302.06590.pdf; Noy, Shakked, and Whitney Zhang (2023) "Experimental evidence on the productivity effects of generative artificial intelligence". SSRN 4375283. https://economics.mit.edu/sites/default/files/inline-files/Noy_Zhang_1_0.pdf.

Figure 3: **Developers with less experience benefit relatively more than more experienced developers**

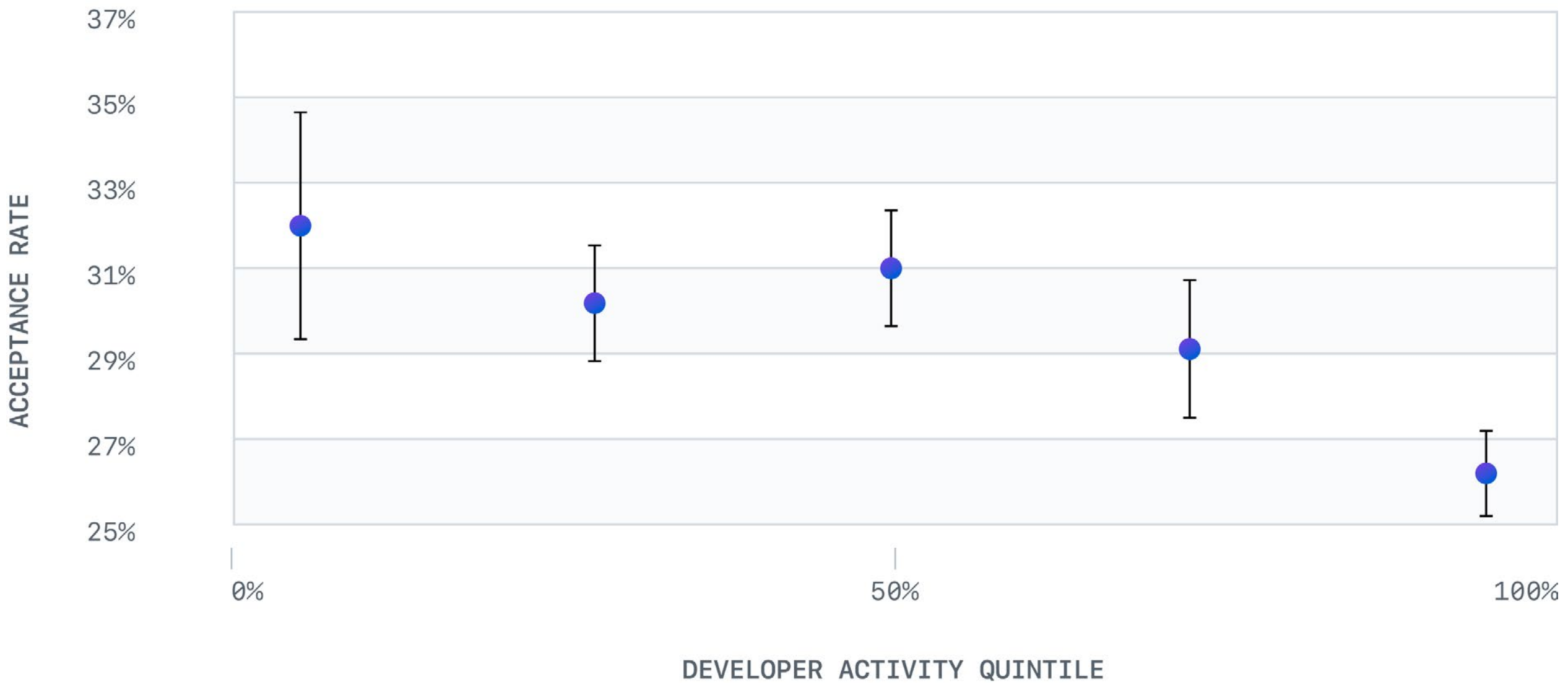

With a sample of nearly one million developers, the data demonstrates GitHub Copilot's potential to substantially impact productivity for the developer community. As developers continue to become fluent in prompting and interacting with AI, particularly with new models that allow natural language to power the development lifecycle, we anticipate that approximately 80% of code will be written with AI. We believe this will help democratize software development, allowing more people from non-technical backgrounds to build and shape the software that will continue to power the global economy.

*These predictions reflect the promise of the technology, but more importantly accelerate the innovative potential of developers.*

The history of software development has seen many new technologies arise, from punch cards to compilers, high-level programming languages, and open source. Developers adopt new tools and use them to benefit themselves and society. Today, there are more people working in software development jobs than ever before and they make more income than ever before, even adjusted for inflation.[14] We expect that as generative AI becomes more widely adopted, the types of tasks required in jobs will also evolve—often to higher-order work.[15] The upper bar on productivity growth from generative AI is difficult to predict because it will also create possibilities for new business models, as witnessed with previous general purpose technologies like the steam engine and the internet.

[14] The US Bureau of Labor Statistics reports that in 1999 there were 2,620,000 people working in Computer and Mathematical Occupations. Their median annual wage was $51,990 ($84,573 in 2021) with an average of $54,930 ($89,355 in 2021). In 2021, there were 4,654,750 people working in these professions, with a median wage of $97,540 and an average of $99,860.

[15] Bessen, James. (2015). "Toil and technology: Innovative technology is displacing workers to new jobs rather than replacing them entirely". Finance & Development 52, no. 001 (2015). https://www.elibrary.imf.org/view/journals/022/0052/001/article-A007-en.xml?rskey=ObbFpr&result=1; Bessen, James E. (2015). "How computer automation affects occupations: Technology, jobs, and skills". Boston Univ. School of Law, Law and Economics research paper 15-49 (2016). https://papers.ssrn.com/sol3/papers.cfm?abstract_id=2690435 & Groshen at al. (2019). "Preparing U.S. Workers and Employers for an Autonomous Vehicle Future". https://research.upjohn.org/cgi/viewcontent.cgi?article=1039&context=up_technicalreports.

# The robust ecosystem of generative AI

The open source community is a critical source of software innovation. In recent years, we have seen an increased focus of the open source community on generative AI foundation models, plugins, and applications, with a diversity of contributions from academia, industry, and individuals. As the home of a large swath of open source community innovation, GitHub is an important platform to enable and measure the sizable opportunity of generative AI.

*As our analysis of GitHub data illustrates, there has been considerable growth in generative AI work on GitHub.*

This increase has expanded from foundations models to codebases for a variety of business applications, spanning multiple layers of the generative AI stack (Figure 4):

**Hardware:** Training on large datasets requires significant hardware investment and computing power, as evidenced by big tech organizations.

**Foundation models:** AI software trained on a large amount of data that can be used to accomplish multiple downstream tasks. These are largely utilized by start-ups, academia and institutions, but big tech is equally present. Some examples include Stable Diffusion and LLaMA.

**Applications:** Today, the three main output modes from generative AI include code, text, and image generation. These are being used in a range of applications, including customer service agents, programming assistants, generation of synthetic data, as well as in gaming and simulation.

Figure 4: **Generative AI Stack**

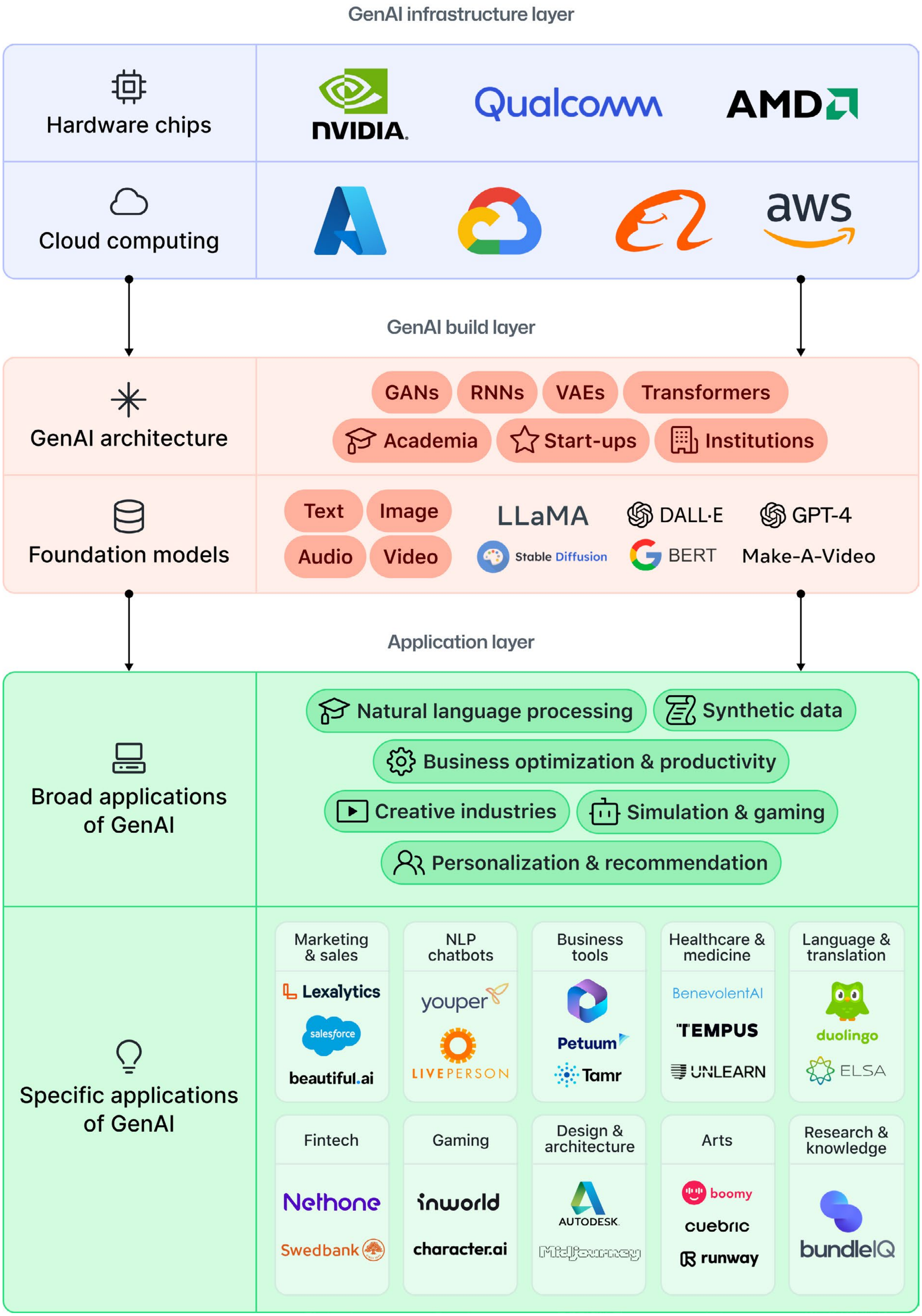

## Public repositories on GitHub

GitHub data shows that open source AI innovation is diverse and is being led by individuals. Analyzing the top 20 account owners—based on stars on their repositories related to generative AI on GitHub[16]—nine belong to individuals and two belong to big tech, and a few others belong to smaller organizations or start-ups that are focused on democratizing AI research, making machine learning models more accessible to the public, providing tools to help deep learning researchers, and working on experimental projects. Of the top 20 account owners by the number of forks on their repositories, 10 belong to individuals. Out of the rest, only two belong to big tech and the others belong to smaller organizations, start-ups, or more than one person. From this data, it's clear that individuals are building with AI on GitHub the most.

## Generative AI is driving a massive surge in developer activity, with the United States leading the way

Figures 5-12 show a spike in growth of models, commits, and forks in repositories related to generative AI.

Figure 5: **Number of generative AI repositories created each week on GitHub**

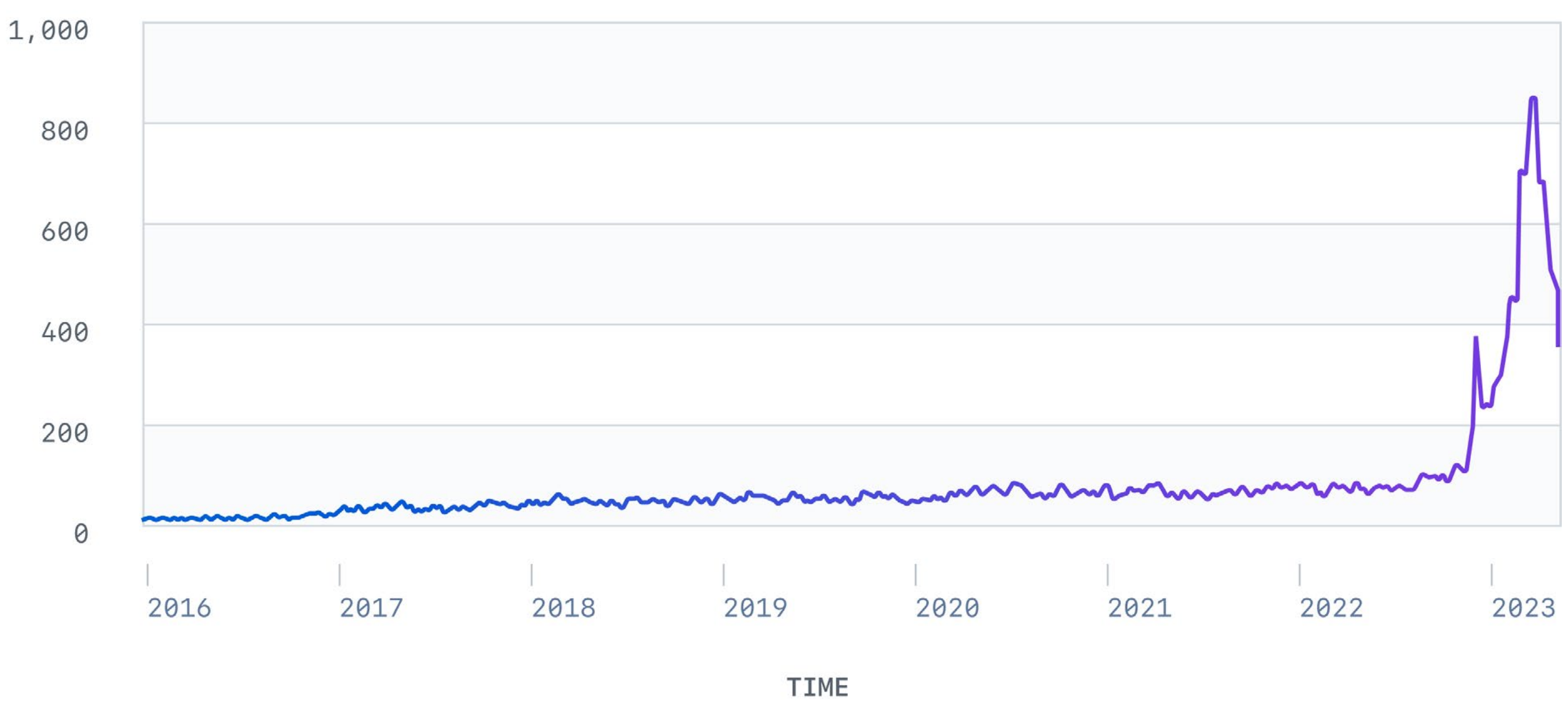

[16] As determined by 683 topics associated with generative AI repositories listed in the Appendix.

Figure 6: **Number of forks on generative AI repositories created each week on GitHub**

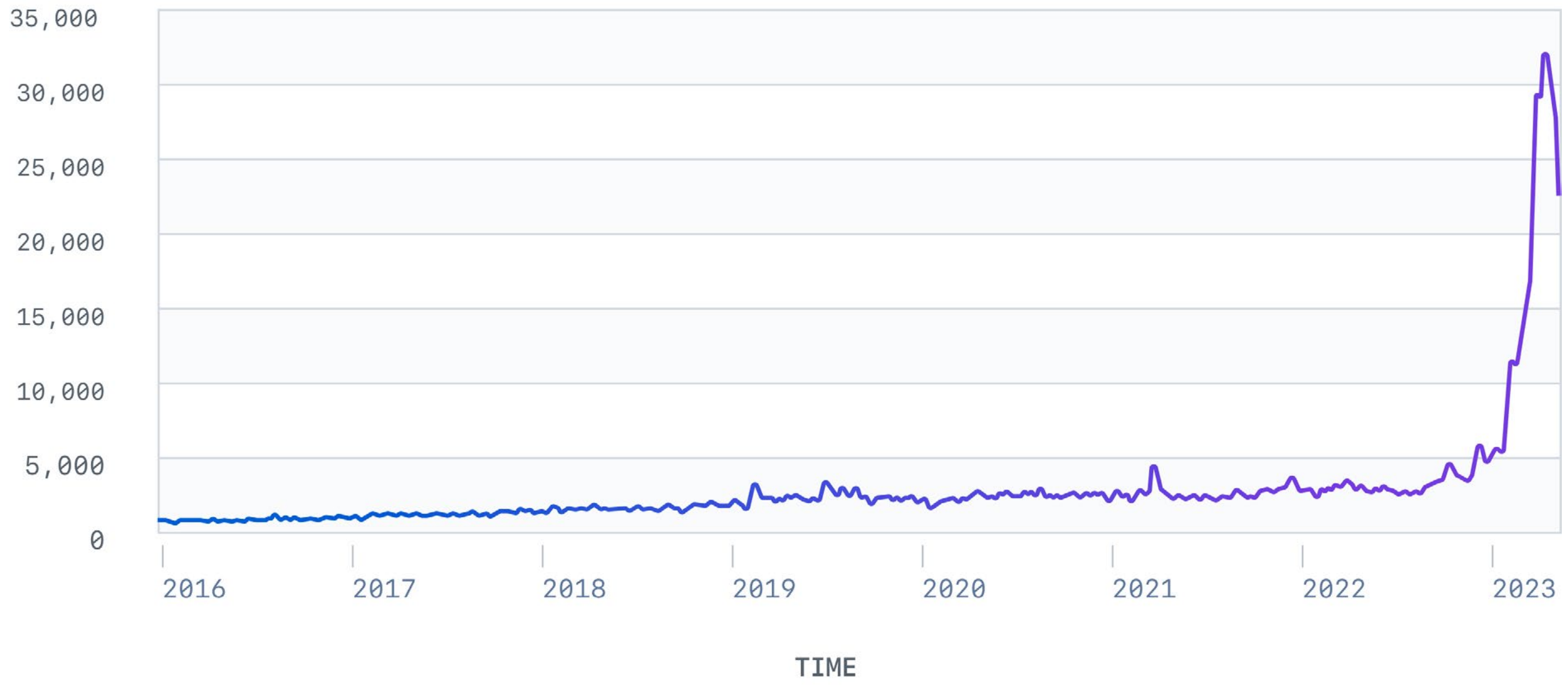

Figure 7: **Monthly growth in the number of commits in generative AI repositories on GitHub**

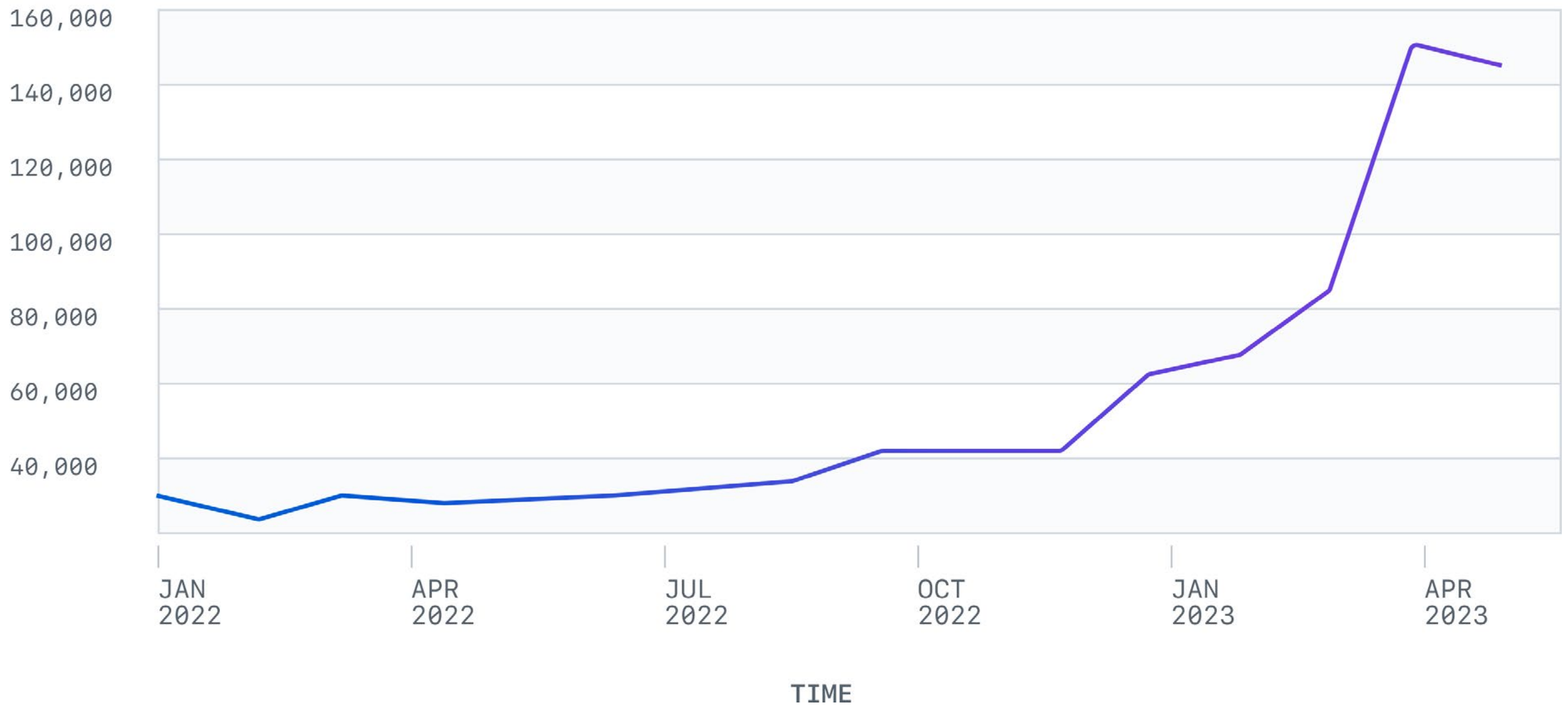

Figure 8: **Monthly growth in the number of contributors in generative AI repositories on GitHub**

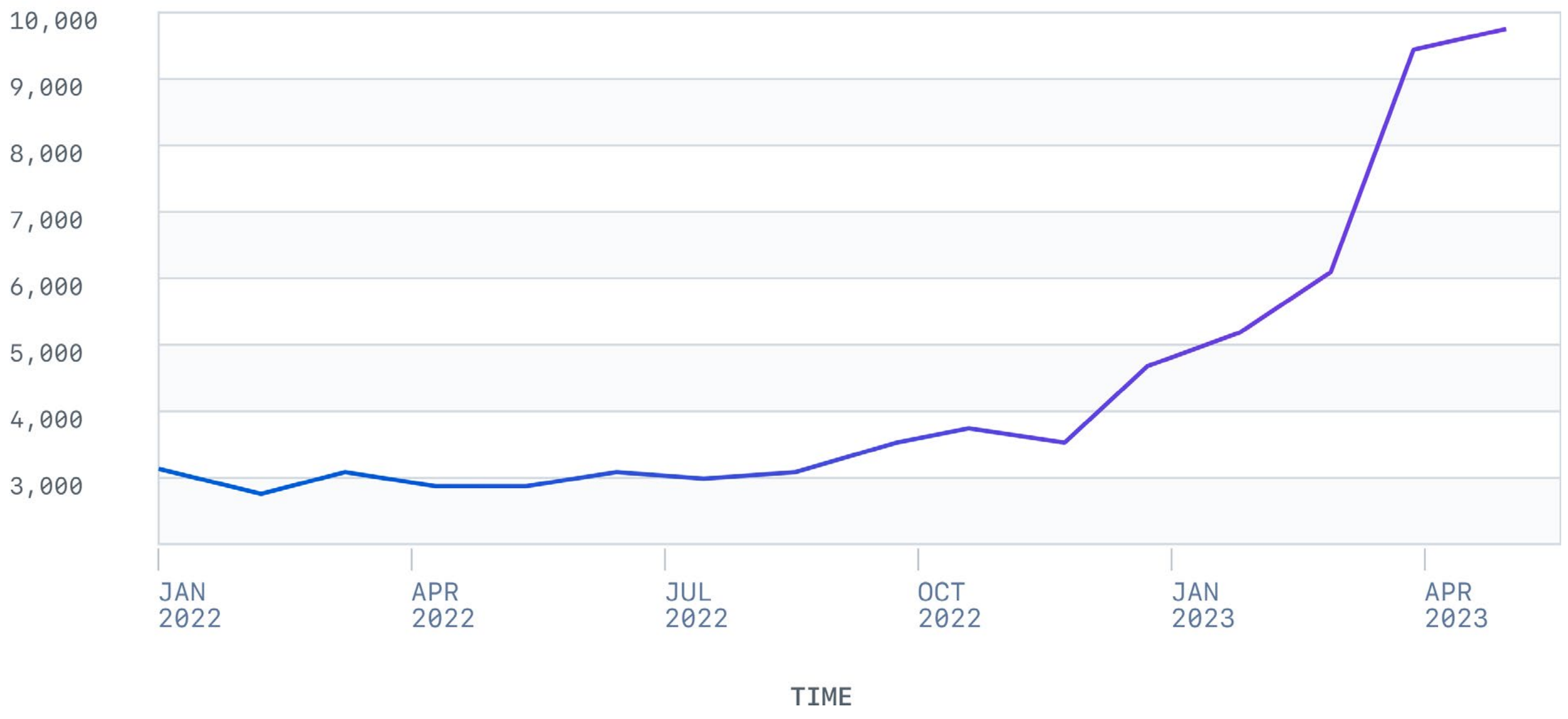

Figure 9: **Number of generative AI repositories created each week on GitHub**

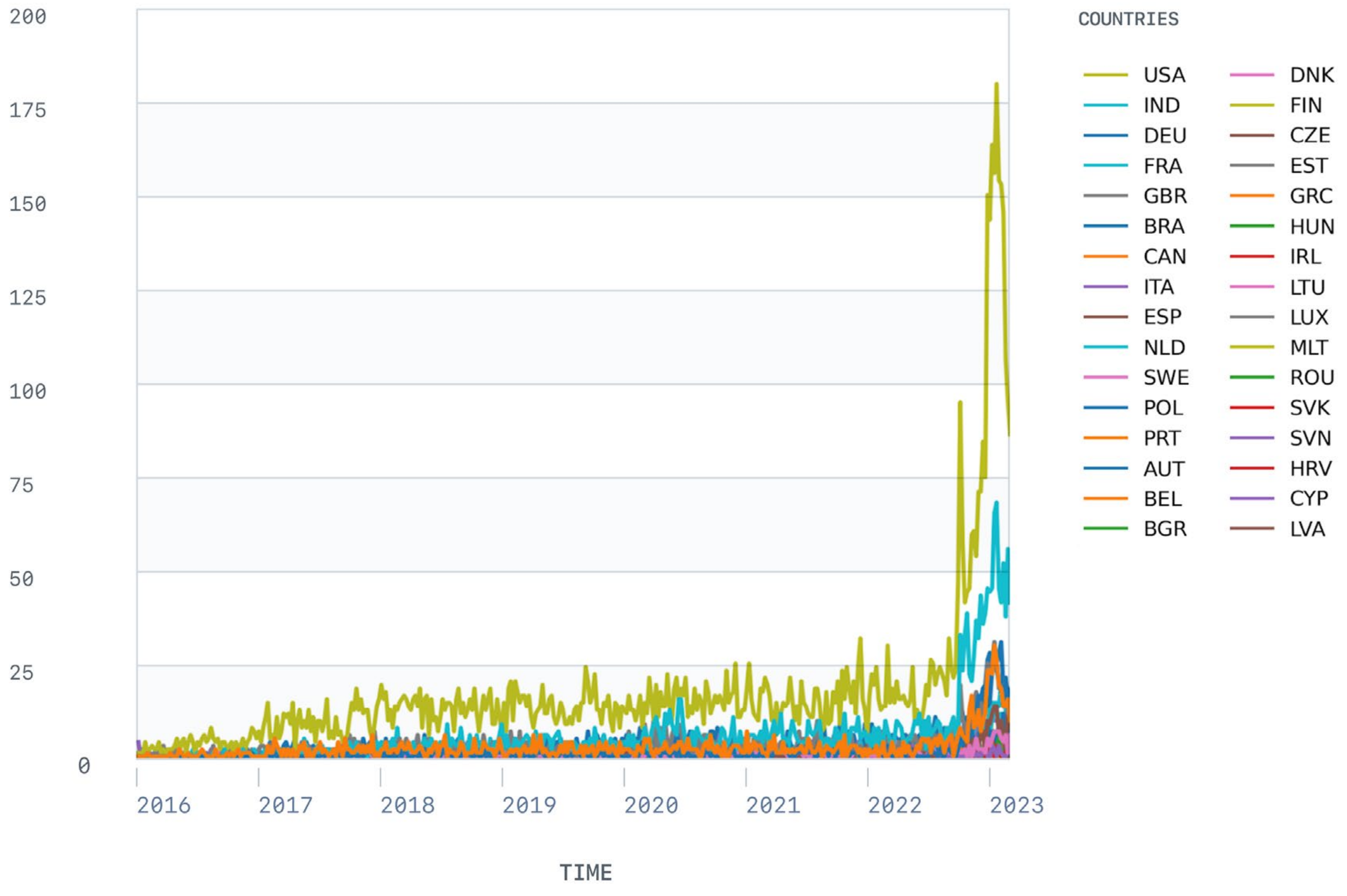

Figure 10: **Number of forks on generative AI repositories created each week on GitHub**

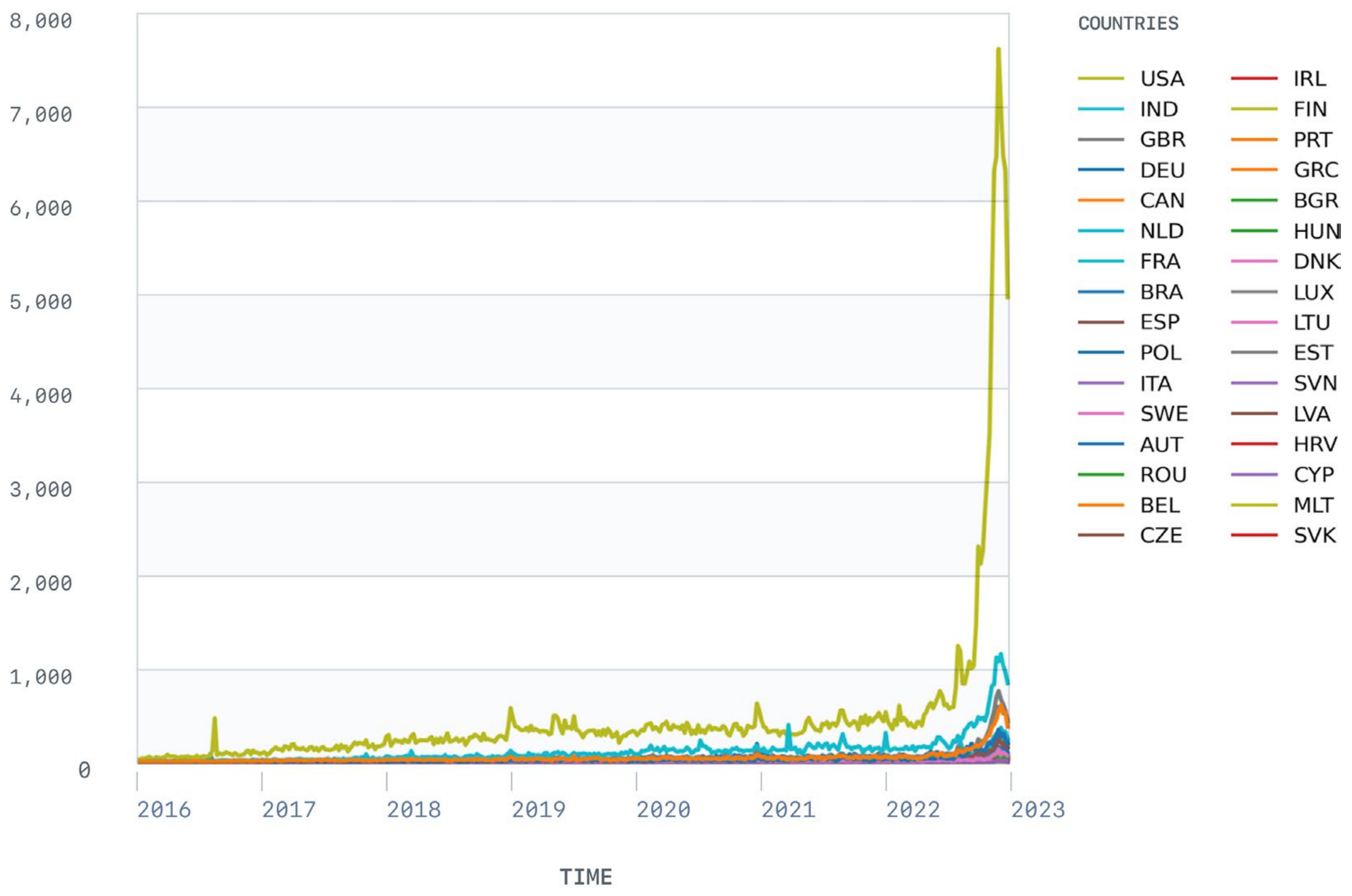

Figure 11: **Monthly growth in the number of commits in generative AI repositories on GitHub**[17]

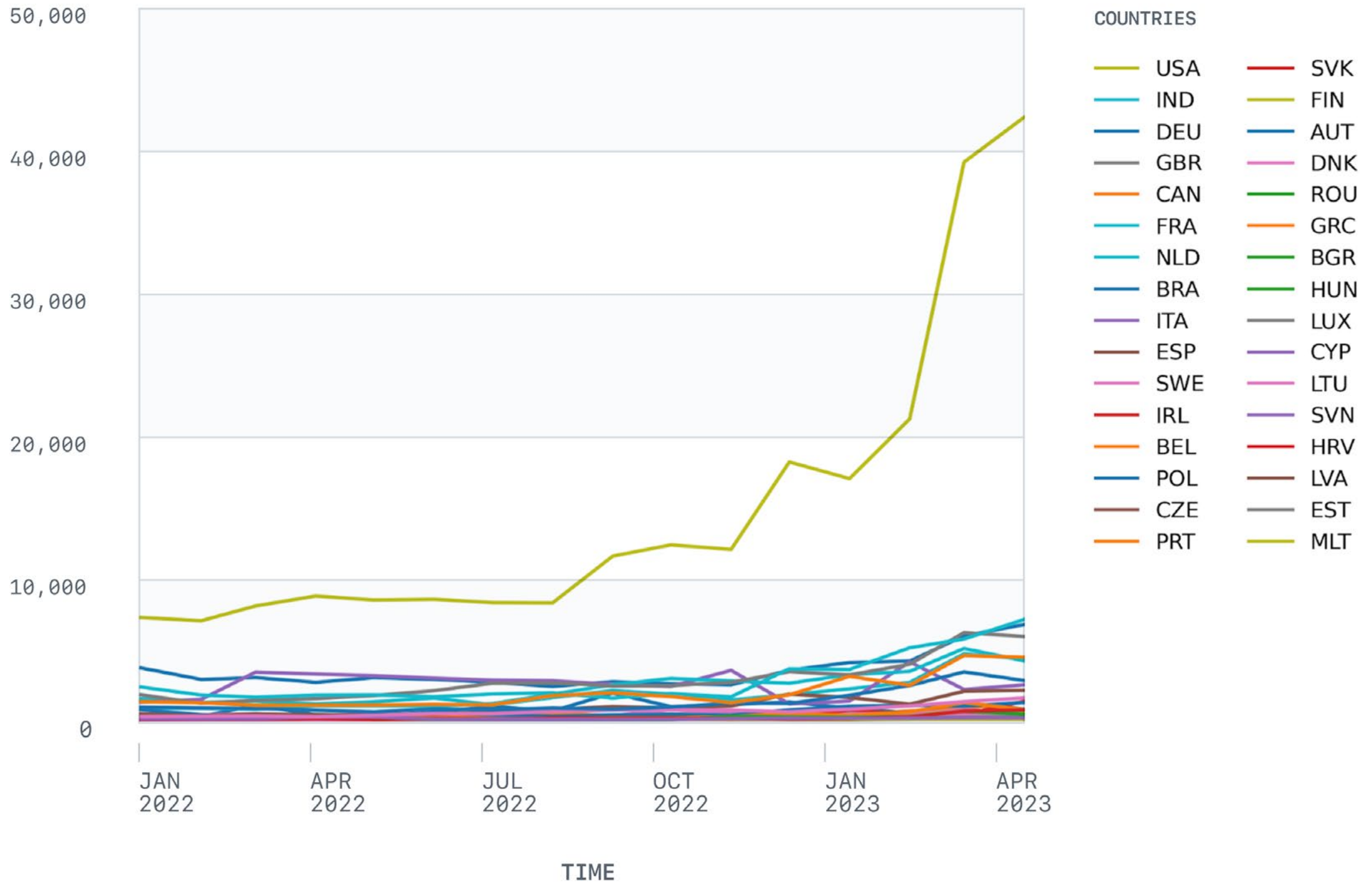

[17] These commits are for generative AI repositories with user reported locations attached to them.

Figure 12: **Monthly growth in the number of contributors in generative AI repositories on GitHub**[18]

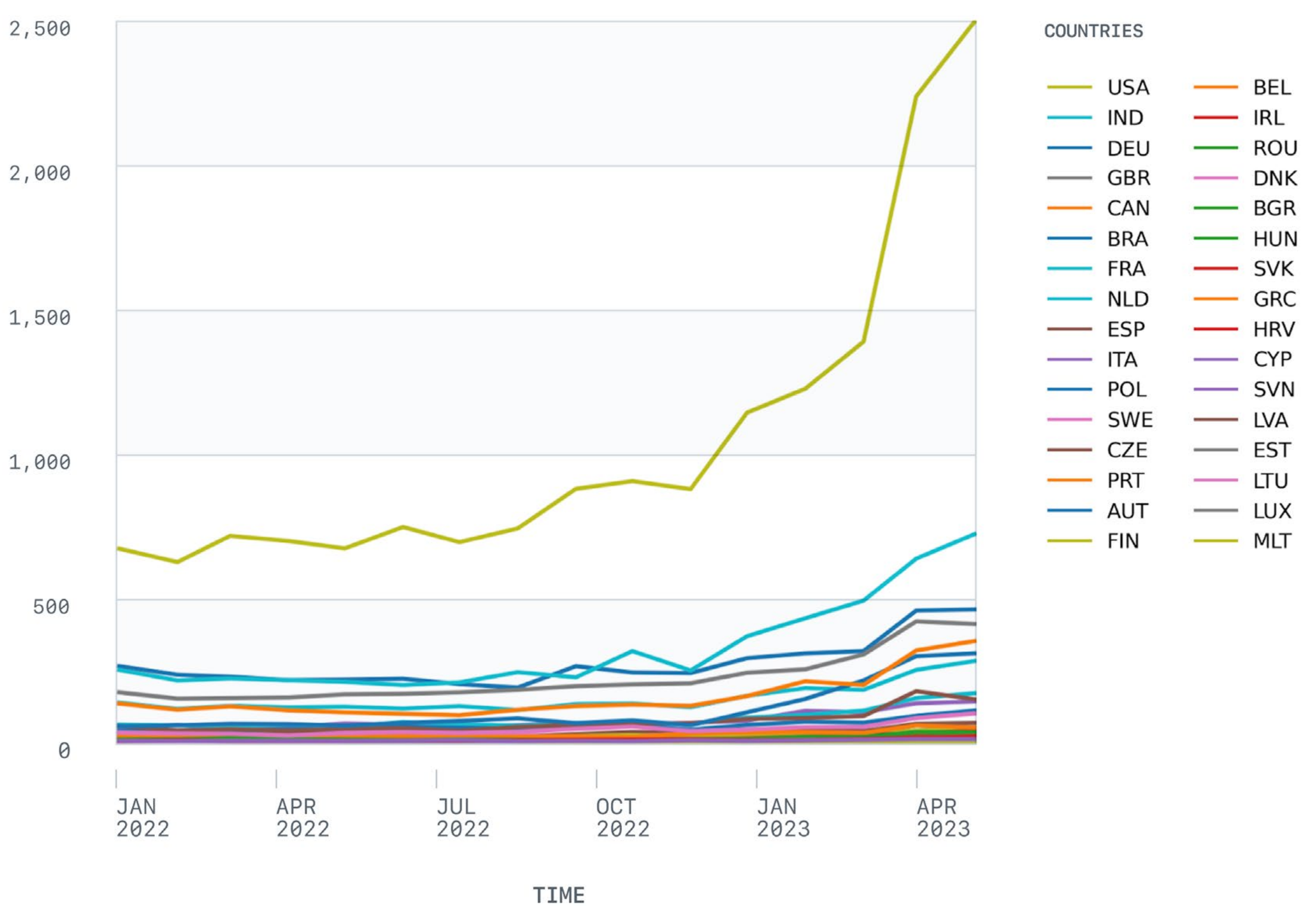

## Ecosystem analysis

After examining the top topics assigned to repositories related to generative AI created each year,[19] we found that the range of projects in generative AI has increased exponentially in recent years, with a tremendous focus on growth in GPT and ChatGPT-related applications (Figure 13).

[18] These commits are for generative AI repositories with user reported locations attached to them.

[19] For identifying repositories related to generative AI, we started with a list of terms related to generative AI, and any particular repository that had any of the terms in their topic label or in the description of the repository was used to classify the repository as related to generative AI. After getting the initial list of repositories, a list of 14,000 topic labels were created from the topics associated with the initial list of repositories and were classified as related to generative AI or not, and the new set of terms were used to query a final list of repositories related to generative AI.

Figure 13: **Top words in the topics on generative AI repositories on GitHub**

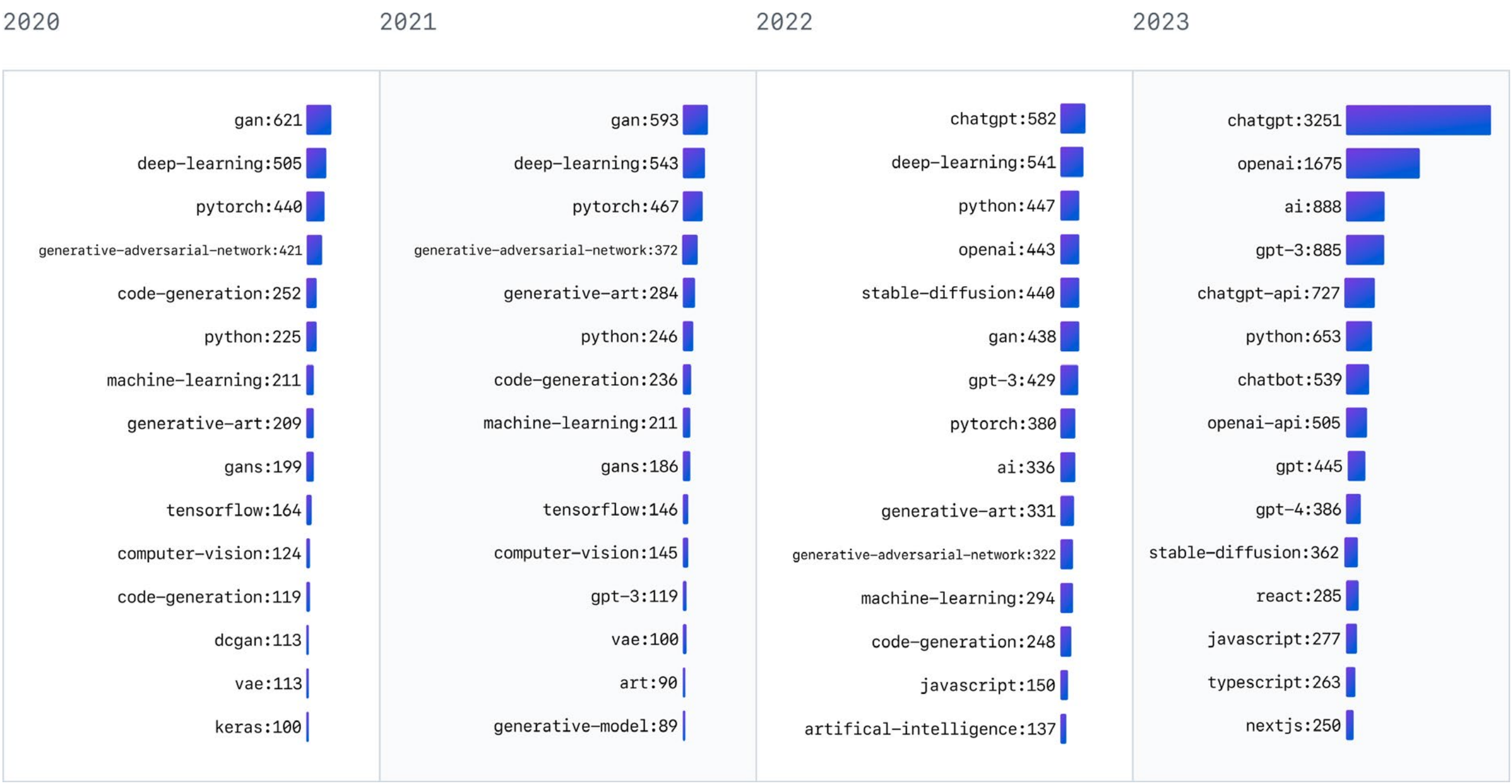

A topic modeling exercise (Figure 14) on the repository descriptions for repositories related to generative AI also showed similar patterns of growth in ChatGPT-related topics.

Figure 14: **Evolution of topics over time**

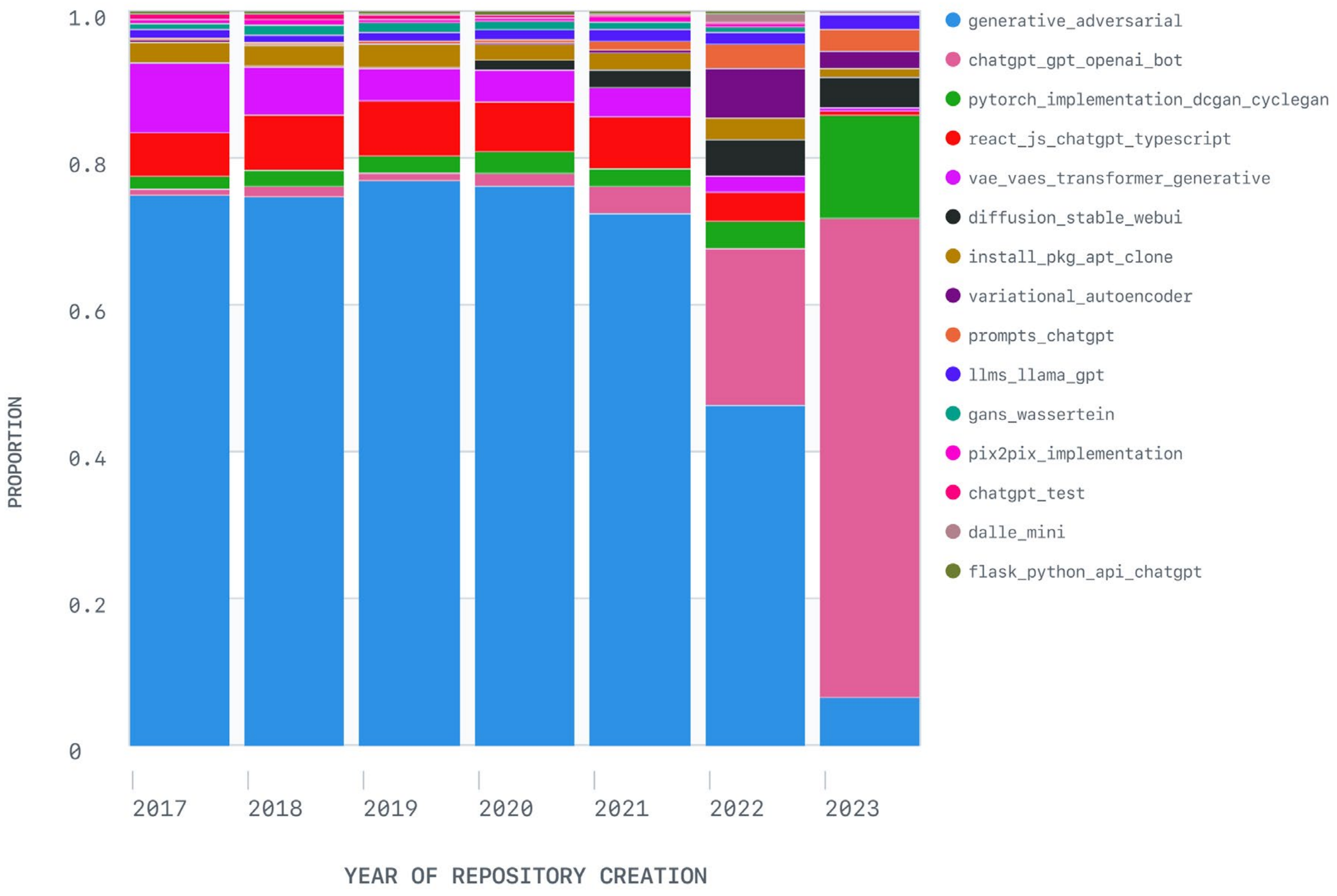

## Diverse development of GPT applications

GPT language models are being integrated with user-facing tools, such as APIs, bots, assistants, mobile applications, and plugins, and are leading to more software development. This has led to a growth of new categories of skills such as prompt-engineering. Open source alternatives to ChatGPT, such as Dolly, as well as other advancements in using ChatGPT, such as alternative web UIs for ChatGPT like ChatGPT Demo and Chatbot-UI have been built off the initial explosion of GPT and ChatGPT. Figure 15 shows a word cloud of topics related to generative AI, indicating the broad set of related or underlying technologies.

Figure 15: **Word cloud of topics for repositories related to ChatGPT on GitHub**

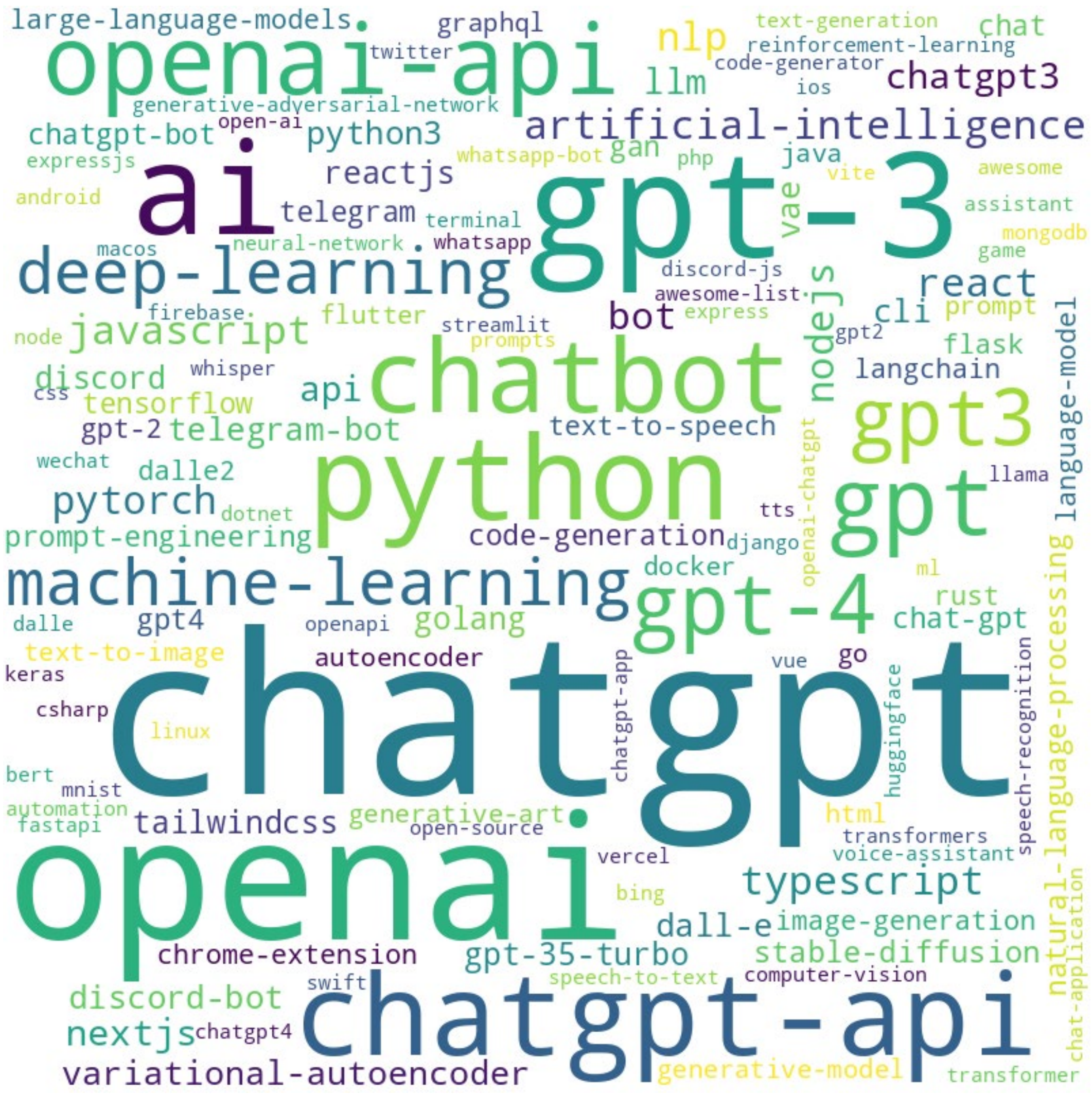

# Economic benefits of generative AI

Technological advancement plays a major role in economic growth.[20] One research study estimated the role of technological innovation on GDP in developing countries during the period 1990–2018 and found technological innovation indicators lead to an increase in economic growth in both the short and long term. In the long run, there is a two-way causal relationship between technological innovation and GDP, and in the short run, causation works unidirectionally, spanning from technological innovation to GDP.[21] Technology can simultaneously benefit workers and capital owners by increasing worker productivity and worker leisure time, as well as GDP.[22]

One line of research established an empirical relationship between open source software and GDP growth by region. A study for the European Commission calculated that open source contributed between €65 EUR and €95 EUR billion to GDP in 2018, and that a 10% increase in contributions would generate between 0.4% and 0.6% additional EU GDP per year. These analyses were performed with GitHub data, and with the benefit of additional data for subsequent years, we can report that commits in the EU grew by 19.7% between 2019 and 2020. This suggests a contribution of between €110 EUR and €164 EUR billion to GDP in 2020 alone. The economic value of open source activity continues to grow, and will only be further accelerated by AI.

---

[20] Robert M. Solow. (1956). "A contribution to the theory of economic growth". The Quarterly Journal of Economics, Vol. 70, No.1. https://pages.nyu.edu/debraj/Courses/Readings/Solow.pdf; Park et al. (2020). Demographic Change, Technological Advances, And Growth". No.617 ADB Economics Working Paper Series. https://www.adb.org/sites/default/files/publication/618796/ewp-617-demographic-change-tech-advances-growth.pdf; Broughel at al. (2019). "Technological Innovation and Economic Growth: A Brief Report on the Evidence". Mercatus Research Paper https://papers.ssrn.com/sol3/papers.cfm?abstract_id=3346495; Niluka et al. (2023). "Trade and economic growth: Does the sophistication of traded goods matter?" https://research-repository.uwa.edu.au/en/publications/trade-and-economic-growth-does-the-sophistication-of-traded-goods.

[21] Mohamed et al. (2022). "Causality between Technological Innovation and Economic Growth: Evidence from the Economies of Developing Countries". https://www.mdpi.com/2071-1050/14/6/3586.

[22] Kleinberg et al. (2017). "Human Decisions and Machine Predictions". https://www.nber.org/papers/w23180; Jason Furman Chairman, Council of Economic Advisers. (2016). "Is This Time Different? The Opportunities and Challenges of Artificial Intelligence". https://obamawhitehouse.archives.gov/sites/default/files/page/files/20161212_cea_nas_ai_furman.pdf; Graetz et al. (2015,). "Robots at work". https://papers.ssrn.com/sol3/papers.cfm?abstract_id=2589780; Executive Office of the President. (2016). "Artificial Intelligence, Automation and the Economy". https://obamawhitehouse.archives.gov/sites/whitehouse.gov/files/documents/Artificial-Intelligence-Automation-Economy.PDF.

Figure 16 below presents different estimates of productivity growth that could be attributed to AI from various academic research papers.[23] These studies estimate the impact of AI, and provide valuable insight into the productivity gains that generative AI will bring for developers.

Figure 16: **Productivity estimates of AI**
Effect of AI Adoption on Annual Worker Productivity Growth Within Firm
*Average: 3.1%, Median: 2.6%*

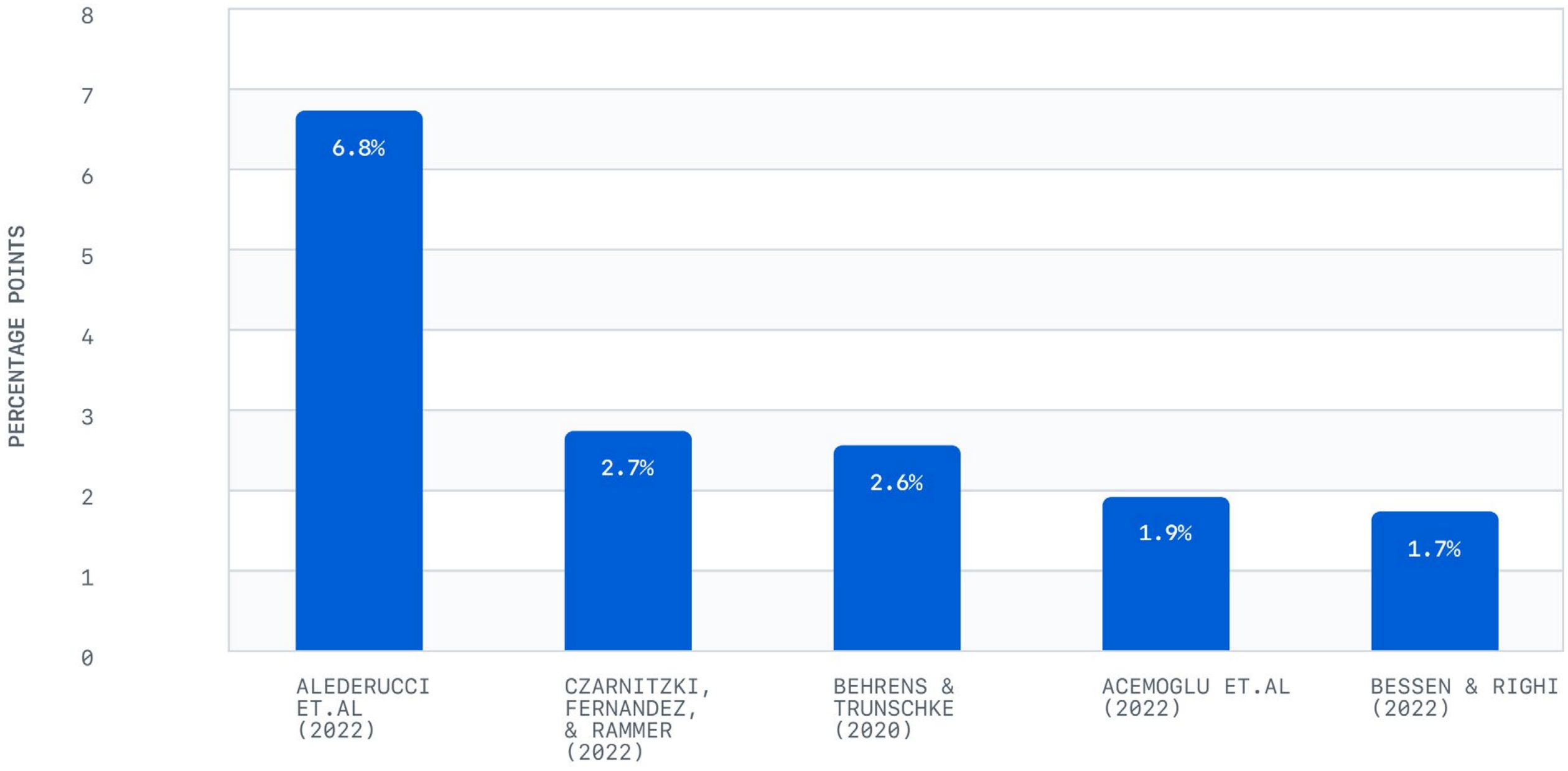

These productivity impacts of AI are only a lower bound of the impact that generative AI will ultimately create. As discussed above, developers are the backbone of productivity today. Research on the effect of immigration, particularly an influx of high-skilled H1B workers with the largest share of those in computer-related occupations in the United States, found that an increase of skilled immigrants lowered the prices and raised the output of IT goods between 1.9% and 2.5%, thus benefiting United States consumers.[24]

Another study evaluated the total factor productivity after the year 2000. Although the IT-producing sector decelerated in the period immediately post 2000 relative to the IT boom, it still accounted for 40% of aggregate productivity growth. The deceleration was counterbalanced by the contribution from industry sectors using IT, which buoyed aggregate Total Factor Productivity (TFP) growth to almost the same rate as the 1995–2000 period.[25]

---

[23] Briggs/Kodnani. (2023). "The Potentially Large Effects of Artificial Intelligence on Economic Growth". Goldman Sachs Economics Research. https://www.gspublishing.com/content/research/en/reports/2023/03/27/d64e052b-0f6e-45d7-967b-d7be35fabd16.html.

[24] Bound et al. (2017). "Understanding the Economic Impact of the H-1B Program on the U.S." https://www.nber.org/papers/w23153.

*There is plenty of evidence showing that the demand in this sector is going to grow over the coming few years in the United States.*

In one study, The Bureau of Labor and Statistics projected the total number of openings in Science & Engineering due to growth, labor force exits, and occupational transfers between 2016 and 2026 to be 6.033 million, including 3.477 million in computer occupations and 1.265 million in engineering occupations.[26] Other projections also show faster than average growth in IT jobs.[27] These are clear signals of growing demand for software and software developers. The productivity boost for developers that GitHub Copilot and generative AI tools provide will significantly aid in seizing this opportunity to meet the accelerating software demand.

It is worth noting that software developers are in demand not only in the traditional software development firms but in a plethora of different industries. With the wave of digitization underway, the demand for software developers is growing at a high rate in a diverse array of industries, such as biotechnology, automotives, manufacturing, data analytics, the financial sector, non-profit sector, and in the government due to transferable skills. As more industries and organizations take a digital approach, the demand for software and software developers will only increase.[28]

---

[25] Jorgenson et al. (2011). "Information technology and U.S. productivity growth: evidence from a prototype industry production account". https://link.springer.com/article/10.1007/s11123-011-0229-z.

[26] "The U.S. Science and Engineering Workforce: Recent, Current, and Projected Employment, Wages, and Unemployment" accessed at https://crsreports.congress.gov/product/pdf/R/R43061/11 on 14th June, 2023.

[27] "The Outlook for In-Demand IT Jobs" accessed at https://blog.dol.gov/2021/09/20/the-outlook-for-in-demand-it-jobs on 14th June, 2023.

[28] "Why Software Developer is the No.1 Job of 2023" accessed at https://money.usnews.com/careers/articles/why-software-developer-is-the-no-1-job-of-2023 on 13th June, 2023.

Globally, there is a growing demand for software and simply not enough developers. Currently there are an estimated 27 million professional software developers in the world.[29,30] In Europe, the shortage in IT jobs is estimated at over 500,000, with Germany exceeding 100,000, and countries like Denmark, Norway, and Sweden expected to face growing demand.[31] Latin America faces a similar shortage, with a deficit of IT workers in Brazil growing at an estimated 25,000 per year.[32] According to an estimate by Korn Ferry, the shortage of tech talent could be 85.2 million in 2030 across various sectors.[33]

*Continued efforts to expand education, employment, and other opportunities to help more people become developers are vital as economies and societies become more reliant on software.*

Generative AI tools hold promise here, not only to skill more developers,[34] but also to make each and every developer more productive.[35] It thus holds promise to meet the exploding demand for software that will only accelerate as AI adoption continues.

[29] "Number of software developers worldwide" accessed at https://www.statista.com/statistics/627312/worldwide-developer-population/ on 13th June, 2023.

[30] The 27 million developer number for 2023 is a conservative number potentially based on the work profile of a person. However, there are other methods to define who a developer is. For example, there were more than 100 million developers on GitHub in January of 2023.

[31] "Solving Scandinavia's talent gap" accessed at https://thescalers.com/the-scandinavian-it-dilemma/ on 13th June, 2023; "Studying IT in Europe: 500.000 developers are missing from the job market" accessed at https://budapestcollege.hu/studying-it-in-europe-500-000-developers-are-missing-from-the-job-market/ on 13th June, 2023.

[32] "Argentina's Digital House raises over $50M to help solve LatAm's tech talent shortage" accessed at https://techcrunch.com/2021/03/25/argentinas-digital-house-raises-over-50m-to-help-solve-latams-tech-talent-shortage on 13th June, 2023.

[33] "The Global Talent Crunch" accessed at https://www.kornferry.com/content/dam/kornferry/docs/article-migration/FOWTalentCrunchFinal_Spring2018.pdf on 14th June, 2023. Although this potentially includes other occupations from different sectors and is a higher bound, this helps us get a sense of the shortage of developers and provides an indication of the critical impact generative AI will create by impacting developer productivity tools.

[34] Recent studies have found that generative AI tools have been especially helpful for new professionals. Peng, Sida, et al. (2023). "The impact of AI on developer productivity: Evidence from GitHub Copilot". arXiv preprint arXiv:2302.06590 (2023). https://arxiv.org/pdf/2302.06590.pdf; Noy, Shakked, and Whitney Zhang. (2023). "Experimental evidence on the productivity effects of generative artificial intelligence." SSRN 4375283 (2023). https://economics.mit.edu/sites/default/files/inline-files/Noy_Zhang_1_0.pdf; Brynjolfsson, Erik, Danielle Li, and Lindsey R. Raymond. (2023). "Generative AI at Work". No. w31161. National Bureau of Economic Research, 2023. https://mitsloan.mit.edu/shared/ods/documents?PublicationDocumentID=9765.

[35] As demonstrated in Section 1.

We established earlier that GitHub Copilot is contributing approximately 30% of code to developers, which increases over time with product improvements and as developers overcome learning curves. This productivity benefit may well be conservative, with other studies finding impacts of 46% and 55%.[36] Using 30% productivity enhancement, with a projected number of 45 million developers in 2030,[37,38] generative AI could add an additional 15 million "effective developers" to worldwide capacity in that timeframe.

*This will have a major impact and help meet accelerating software demand, contributing as much as $1.5 trillion USD to the global economy.*

Even with this boost, demand for software is expected to continue outstripping supply, as demand continues to grow with the spread of digitization across industries. This implies that the additional 15 million developer capacity increase would be fully absorbed by the global economy, with developers remaining in high demand. Consistent with our assumptions, a Korn Ferry report estimates that there will be a shortage of 85.2 million jobs across skilled workers, amounting to a loss of $8.4 trillion USD to global GDP by 2030.[39] The report therefore suggests that each missing skilled worker

[36] "Westpac sees 46 percent productivity gain from AI coding experiment", accessed at https://www.itnews.com.au/news/westpac-sees-46-percent-productivity-gain-from-ai-coding-experiment-596423 23rd June, 2023; Peng, Sida, et al. (2023). "The impact of AI on developer productivity: Evidence from GitHub Copilot." arXiv preprint arXiv:2302.06590 (2023). https://arxiv.org/pdf/2302.06590.pdf.

[37] "The Global Developer Population 2019" accessed at https://slashdata-website-cms.s3.amazonaws.com/sample_reports/EiWEyM5bfZe1Kug_.pdf on 13th June, 2023.

[38] A McKinsey report entitled "JOBS LOST, JOBS GAINED: WORKFORCE TRANSITIONS IN A TIME OF AUTOMATION" (2017, Dec) finds that jobs related to developing and deploying new technologies may also grow, and they estimate that by 2030 this could create 20 to 50 million jobs globally. This prediction bolsters the claim that there would be at a minimum 45 million developers in 2030, given that the current number of developers is 27 million.

[39] "The Global Talent Crunch" accessed at https://www.kornferry.com/content/dam/kornferry/docs/article-migration/FOWTalentCrunchFinal_Spring2018.pdf on 15th June, 2023. The study measures the talent shortage as the gap between supply and demand. They model each industry's share of the economy, taking into account factors such as per-capita income, export position, and natural resource endowments, using OECD long-term forecasts for some of these correlations. To establish the size of the economy itself, they forecast the compound annual growth rate (CAGR) based on the 2000 to 2015 CAGR's relationship with world trade growth, starting level of GDP per capita, growth in government expenditure, and oil rents per GDP. They forecast the educational level of the demanded workforce using known data that we project into the future. To establish labor supply, they use ILO projections of the available labor force to 2030. They derive the number of required workers by dividing the sector's value-added output by the productivity rate (the value added per worker). To distribute the available labor force to the three sectors, they allocate available labor at each skill level to the proportionate need by sector. The labor-demand share of each sector is the share of available labor force allocated to each sector, and the talent gap arises from shortfalls.

could be responsible for nearly $100,000 USD in GDP loss by 2030. Assuming that the average developer's impact is similar to or higher than the impact of a skilled worker (which also seems quite conservative), having 15 million additional developers could then translate to an increase of as much as $1.5 trillion USD.

## Generative AI has seen a spike in funding growth

The massive value that generative AI can bring has been recognized and supported by significant venture capital. Crunchbase, a platform that aggregates data on start-ups, investors, and others, provides a lens into the significant economic activity on generative AI.

The analysis of Crunchbase data shows massive funding activity in the space in recent years.[40] The first figure contains both AI and generative AI firms, since many firms that focus on generative AI do not necessarily classify themselves as generative AI firms on Crunchbase. Since the figure captures funding activity from 2022 onwards, it is reasonable to believe that a substantial portion of funding for AI firms is related to generative AI.

Figure 17: **Time series on funding amounts received by companies classified as AI or generative AI companies on Crunchbase**[41]

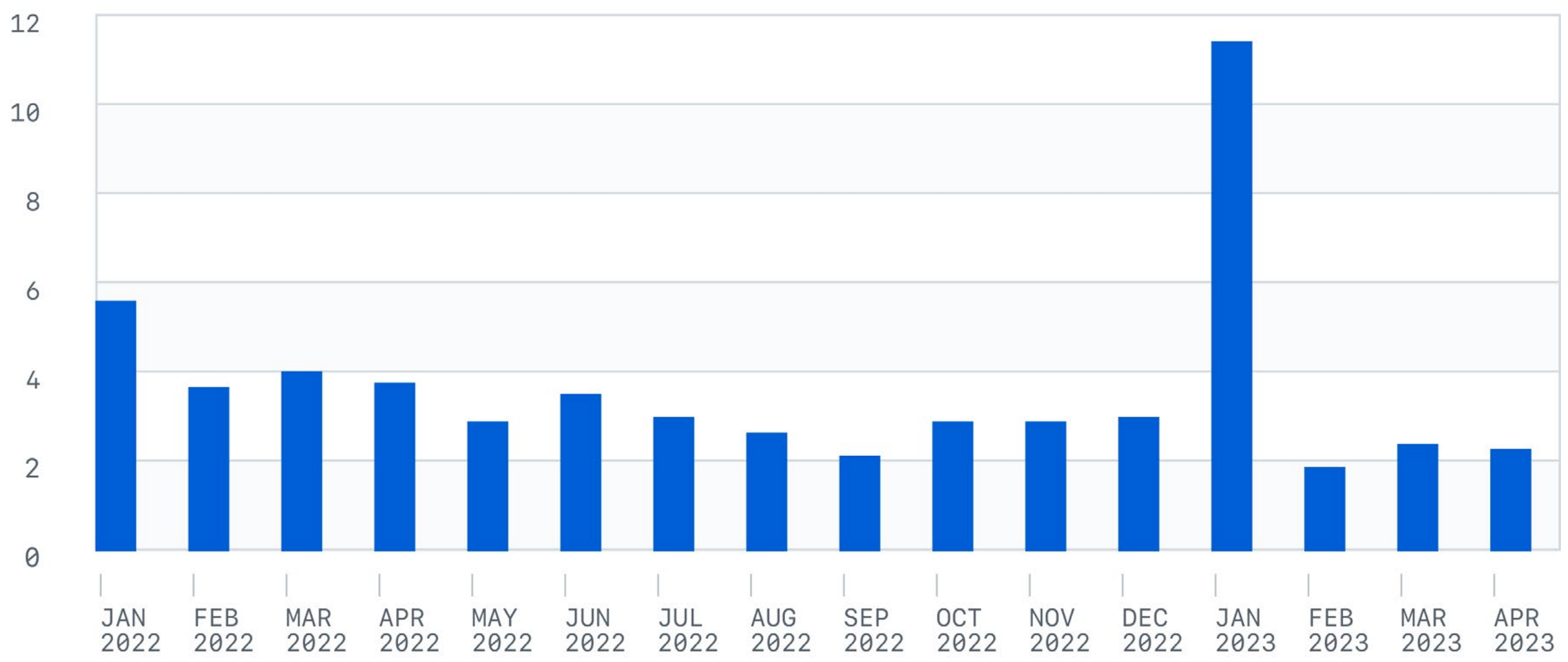

[40] We searched Crunchbase for firms related to AI and generative AI and scraped financial information of all firms who had a defined founding date.

[41] It is worth noting that this dataset shows a spike in funding in January due to the increase in companies founded at the beginning of the year.

The next figure demonstrates the growth in funding for companies that are related to generative AI only.

Figure 18: **Time series on funding amounts received by companies classified as generative AI companies on Crunchbase**

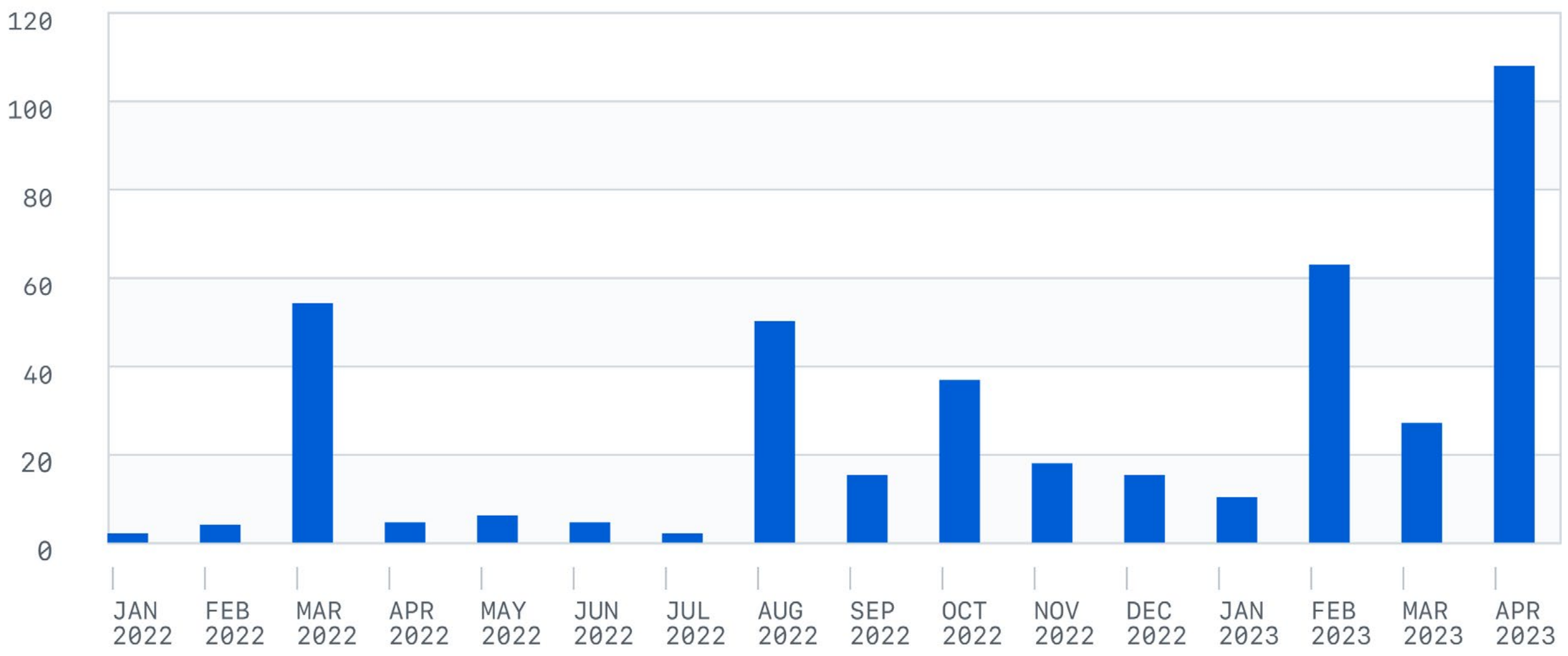

## An inextricable link

From clear gains in productivity, to improved acceptance rates as time goes on, to the explosion of generative AI in open source repositories, to the potential 1.5 trillion to the global GDP, these findings signify a sea change in software development. As more developers adopt these tools and become fluent in the skill set of prompting with generative AI, it is clear that this new way of software development has created an inextricable link between humankind and artificial intelligence that could well define how the world's software is built for generations to come. We hope this paper invites further exploration and research on the productivity improvements and economic impact of the generative AI–powered developer lifecycle.

*The authors are thankful to Demetris Cheatham, Peter Cihon, Cole Driver, Jobey Greenwood, Mike Linksvayer, Julia Neagu, Sam Oshin, Preston Price, Shonte Stephenson, Cory Thayer, and Kevin Xu from GitHub, Sida Peng from Microsoft, and Shilpi Mukherjee, James Palano, and Bartley Tablante from Keystone for providing support with data, analysis and for their invaluable feedback. The authors are also thankful to Professor Blind Knut from Technical University of Berlin and Fraunhofer ISI for support with data requests.*

# Appendix

## Learning Effects hold for Total Acceptances made by developers

We witness a strong effect on learning, by measuring Total Accepts in addition to acceptance rates as shown below controlling for the total number of recommendations. The table shows the coefficients and the p-values. The const denotes the constant term in the regression. We see a statistically significant positive coefficient of 0.0491 for the main variable of interest DAYS_SINCE_USER_START (the number of days since a user first started on GitHub Copilot) while controlling for the total number of recommendations shown to the user, denoted by TTL_SHOWS.

| | coef | std.err | t | P>\|t\| | [0.025 | 0.975] |
|---|---|---|---|---|---|---|
| const | -6.6039 | 0.010 | -629.966 | 0.000 | -6.624 | -6.583 |
| DAYS_SINCE_USER_START | 0.0491 | 0.000 | 364.890 | 0.000 | 0.049 | 0.049 |
| TTL_ SHOWS | 0.3530 | 3.35e-05 | 1.05e+04 | 0.000 | 0.353 | 0.353 |

The results also hold if we add individual fixed effects, filter for developers who have more than 10 recommendations a day, or if we account for different countries.

# Generative AI projects on GitHub

The following list of 683 terms were used to classify a repository on GitHub related to generative AI:

| | | |
|---|---|---|
| 3d-gan | ebgan | pokemon-gan |
| 3d-gans | edgegpt | pose-gan |
| 3d-iwgan | eigengan | promptengineering |
| 3dmm | engan | prompt-engineering |
| 3d-sdn | enlightengan | pre-trained-gan |
| 3d-vae | esrgan | progan |
| adversarial-autoencoders | evolutionary-art | progressive-gan |
| adversarial-filtering | evolutionary-gan | progressive-growing-of-gans |
| adversarial-machine-learning | face-gan | progressively-growing-gan |
| adversarial-network | fastsrgan | psgan |
| adversarial-networks | fcfgan | pytorch-cgan |
| adversarial-perturbations | fgan | pytorch-cgan-mnist |
| adversarial-training | f-gan | pytorch-conditional-gan |
| 3d-generation | fishergan | pytorch-dcgan |
| a2c | fisher-gan | pytorch-gan |
| a2c-algorithm | folk-rnn | pytorch-gan-cgan |
| a3c | gan | quantum-gan |
| a3c-lstm | freechatgpt | quaternion-gan |
| aae | gans | ragan |
| acgan | full-band-melgan | ralsgan |
| ac-gan | general-adversarial-network | ranking-cgans |
| albert-transformer | generated-novel | rasengan |
| advgan | generate-images | rasgan |
| aggan | generation-algorithms | real-esrgan |
| ai-generated-images | generative-adversarial-net | realesrganv2-gu |
| ai-generator | generative-adversarial-nets | recurrent-gan |
| ai-image-generation | generative-adversarial-network | relativistic-gan |
| ai-music-generator | generative-adversarial-networks | residual-cycle-gan |
| aitextgen | generative-adversarial-transformer | revchatgpt |
| ai-text-generator | generativeai | revgan |
| ai-text-to-image-generation | generative-ai | rsgan |
| aivideocreation | generative-ai-pharmacist | sagan |
| algorave | generative-algorithm | sa-gan |
| alias-free-gan | gamma-vae | sarigan |
| animegan | gan-pix2pix | scorchai |
| anime-gan | generating-faces | score-based-generative-modeling |
| animegan2 | generative | score-based-generative-models |
| animeganv2 | generativeart | segan |
| anogan | generative-art | se-gan |
| anogan-keras | generative-audio | selectiongan |
| apdrawinggan | generative-deep-learning | semantic-image-inpainting |
| arcanegan | generative-learning | semi-supervised-gan |
| artgan | generative-literature | seqgan |
| art-generator | generativemodel | seqgan-tensorflow |
| attgan | generative-model | sgan |
| attngan | generative-modeling | simgan |
| autogan | generative-modelling | simple-gan |
| automatic-music-generation | generative-models | simplet5 |
| auxiliary-classifier-gan | generative-music | simpsons-dcgan |
| awesome-chatgpt | generative-neural-network | simsiam |
| awesome-chatgpt-prompts | generative-search | singan |
| began-pytorch | generative-testing | sin-gan |
| beta-tcvae | generative-text | sngan |
| beta-vae | generattive | sngan-projection |
| betavae-metric | geneva-gan | sobolev-gan |
| bgan | gfpgan | social-gan |
| bicyclegan | ggan | soft-introvae |
| bigan | gitagpt | soft-intro-vae |
| biggan | gpt | softmax-gan |
| biggan-encoder | gpt- | spatial-attention-gan |
| biggan-models | gpt2 | stability-ai |
| bing-chat | gpt-2 | stablediffusion |
| binggpt | gpt2-piano | stable-diffusion |
| blended-diffusion | gpt-2-simple | stable-diffusion-api |
| boundary-equlibrium-gan | gpt-2-text-generation | stable-diffusion-diffusers |
| brain-gnn | gpt-2-tokenizer | stable-diffusion-embedding |
| bsrgan | gpt3 | stable-diffusion-library |
| calogan | gpt-3 | stable-diffusion-tutorial |
| camstyle | gpt35 | stable-diffusion-ui |
| cartoongan | gpt-35 | stable-diffusion-v1-5 |
| cartoon-stylegan | gpt-3-5 | stable-diffusion-v2 |
| chatglm | gpt-35-burbo | stable-diffusion-webui |
| chatgpt | gpt-35-prompts | stable-diffusion-web-ui |
| chat-gpt | gpt-35-turbo | stable-diffusion-webui-plugin |
| chatgpt3 | gpt-3-5-turbo | spheregan |
| chat-gpt3 | gpt-35-turbo-0301 | srgan |
| chat-gpt-3 | gpt3-cli | srgan-pytorch |
| chatgpt35-turbo | gpt3-davinci | ssgan |
| chatgpt4 | gpt3-decoder | stackgan |
| chatgpt-4 | gpt3-email | stack-gan |
| chatgpt-android | gpt3-encoder | stargan |
| chatgpt-answer | gpt3-instruct | stargan-tensorflow |
| chatgpt-api | gpt3-library | stargan-v2 |
| chatgpt-api-sdk | gpt3-mail | stargan-vc |
| chatgpt-api-wrapper | gpt-3-prompt | steganalysis |
| chatgpt-app | gpt3-prompts | steganography |
| chat-gpt-app | gpt-3-prompts | storygan |
| chatgpt-awesome | gpt-3-text-generation | structural-gan |
| chatgpt-bot | gpt-3-tokenizer | styled-based-gan |
| chatgpt-browser-extension | gpt3-turbo | stylegan |
| chatgpt-chat-completion | gpt4 | style-gan |
| chatgpt-chat-history | gpt-4 | stylegan2 |
| chatgpt-chrome-extension | gpt-api | stylegan2-ada |
| catgan | gptchat | stylegan2-ada-pytorch |
| ccgan | gpt-chatbot | stylegan2-model |
| ccgans | gpt-cli | stylegan2-paper |
| cdcgan | gpt-client | stylegan2-pytorch |
| cdcgan-model | gptdetector | stylegan3 |
| cgan | gpt-detector | stylegancpp |
| cgan-mnist | gpt-index | stylegan-encoder |
| cgans | gpt-models | stylegan-image-manipulation |
| cgan-training | gpt-neo | stylegan-inversion |
| chatgpt-cli | gpt-neo-fine-tuning | stylegan-model |
| chatgpt-client | gpt-neo-hugging-face | stylegan-pytorch |
| chatgptclone | gpt-neo-text-generation | style-melgan |
| chatgpt-clone | gpt-neox | sylegan |
| chatgpt-code-generation | gpt-neo-xl | tacotron |
| chatgpt-dart | gpt-priming | tacotron2 |
| chatgpt-detector | gpts | tacotron-2 |
| chatgptdiscord | glow-tts | tacotron2-pytorch |
| chatgpt-discord | google-chatgpt | tacotron-pytorch |
| chatgpt-discord-bot | gp-gan | text-davinci |
| chatgpt-firefox-extension | gpt3-resources | text-davinci-002-render |
| chatgpt-flutter | gpt3-terminal | text-davinci-002-render-sha |
| chatgpt-go | gptj | text-davinci-003 |
| chatgpt-i18n | gpt-j | text-from-image |
| chatgpt-image-generation | gpt-j-6b | textgeneration |
| chatgpt-information-retrieval | gpt-script | text-generation |
| chatgpt-in-ms-word | gpt-sdk | text-generation-webui |
| chatgpt-in-word | gpt-terminal | text-generation-with-prompting |
| chatgpt-io | gpt-turbo | text-generator |
| chatgpt-ir | gpt-voice | tensorflow-gan |
| chatgpt-java | gptzero | tensorflow-pix2pix |
| chatgpt-jetbrains | hackgpt | textgenrnn |
| chatgpt-output | hayao-style | texttoimage |
| chatgpt-plus | hologan | text-to-image |
| chatgpt-prompts | hypergan | text-to-image-ai |
| chatgpt-python | hypergraphs | text-to-image-generation |
| chatgpt-sdk | image-generation-model | text-to-image-search |
| chatgpt-text-completion | image-generator | text-to-image-synthesis |
| chat-gpt-tool | image-generators | tf2gan |
| chatgpt-turbo | image-generator-using-openai-api | tfgan |
| chatgpt-tutorial | image-gpt | tl-gan |
| chatgpt-web | information-retrieval-chatgpt | two-player-gans |
| chatgpt-whatsapp-bot | instructgpt | txt2img |
| chatgtp | instruct-gpt | txt2img-generation |
| ciphergan | interactive-generative-art | txt-to-img |
| code2seq | interactive-image-generation | txt-to-img-generation |
| codegen | interactive-text-generation | training-gans |
| codegeneration | ir-chatgpt | transferring-gans |
| code-generation | list-of-algorithms-used-in-generative-ai | transgan |
| code-generator | luna-diffusion | trillion-parameters |
| codex | medical-image-generation | vae |
| codex-editor | melgan | vae-cnn |
| cogan | mel-gan | vae-gan |
| conditional-gans | melgan-stft | vae-model |
| condgan | midi-generation | vae-pytorch |
| conditional-dcgan | midigenerator | vaes |
| conditionalgan | midi-generator | u2net |
| conditional-gan | midi-music | u-2-net |
| conditional-image-generation | midi-vae | variationalautoencoder |
| conditional-vae | min-dalle | variational-autoencoder |
| conditional-variational-autoencoder | mingpt | variational-auto-encoder |
| conditional-wasserstein-gan | msg-gan | variational-autoencoders |
| conditonal-gan | multiband-gan | ugan |
| condtional-gan | multiband-melgan | ugatit |
| consingan | multi-band-melgan | unet-gan |
| convert-text-to-image | munit | vaegan |
| cramer-gan | natural-language-generation | vae-gmm |
| create-image-from-text | neuralradiance-fields | vae-implementation |
| create-slide-from-text | neural-radiance-fields | vae-infer |
| creative-ai | neuralrendering | vae-mlp |
| crnngan | neural-rendering | vae-tacotron |
| crnn-gan | neural-scene-representations | vanilla-gan |
| c-rnn-gan | neural-sparql-machines | varational-autoencoder |
| cvae | neuralstyletransfer | v-diffusion |
| cwgan-gp | neural-style-transfer | vd-vae |
| cyclegan | neural-style-transfer-tensorflow | vespcn |
| cycle-gan | neural-text-generation | vid2img2img2vid |
| cyclegan-keras | node-chatgpt | vid2vid |
| cyclegan-pytorch | nuxt-chatgpt | video-inpainting |
| cyclegan-qp | openai-chatgpt | visualgpt |
| cyclegan-tensorflow | open-ai-chatgpt | vit-vqgan |
| cyclegan-vc | openai-codex | vocgan |
| cycleganvc2 | openai-gpt | voice2face |
| cyclegan-vc2 | openai-gpt2 | voice-conversion-gan |
| cyclegan-vc3 | openai-gpt3 | voice-melody-transcription |
| dalemini | openai-whisper | voice-style-transfer |
| dal-e-mini | nvidia-stylegan2 | voice-synthesis |
| dalle | ocr-vqgan | voila |
| dalle- | ogan | vqgan |
| dall-e | openai-app-swift | vqgan-clip |
| dalle2 | openai-clip | vqvae |
| dalle-2 | openai-code-generation | vq-vae |
| dall-e2 | openaiswift | vq-vae2 |
| dall-e-api | openai-swift | vq-vae-2 |
| dall-e-clone | open-ai-transformer | vq-vae-wavenet |
| dall-e-community | openapi-generator | vrae |
| dalle-e | open-generative-ai | vue-chatgpt |
| dallegenerator | outpainting | wae |
| dalle-images | p2x8c4m64p | wae-gan |
| dalle-mega | pacgan | wae-mmd |
| dalle-min | pa-gan | waifu-diffusion |
| dalle-mini | parallel-tacotron | waifu-generation |
| data-efficient-gan-training | parallel-tacotron2 | wasserstein-autoencoder |
| dcgan | parallelwavegan | wasserstein-autoencoders |
| dc-gan | parallel-wavegan | wasserstein-distance |
| dcgan-keras | parallel-wavenet | wasserstein-gan |
| dcgan-mnist-tutorial | parametric-art-generator | wasserstein-gans |
| dcgan-model | patchgan | wav2lip |
| dcgan-pytorch | patch-gan | wav2vec2 |
| dcgans | pathology-gan | waveform-generation |
| dcgan-tensorflow | pix2pix | wavegan |
| deblurgan | pix2pix-autoencoder | wavegan-pytorch |
| deep-biggan | pix2pix-gan-framework | waveglow |
| deepdream | pix2pix-pytorch | wavenet |
| deep-dream | pix2pix-tensorflow | wavenet-keras |
| deep-generative-model | persian-log2vis | wavenet-vocoder |
| deep-generative-modelling | personagpt | wavernn |
| deep-generative-models | pggan | wavevae |
| dfc-vae | pggan-tensorflow | wavlm |
| dialogpt | pix2p | wgan |
| dialogue-generation | pix2pix3d | wgan-cp |
| dialogue-generator | pix2pixhd | wgangp |
| diffgan-tts | pix2pix-m | wgan-gp |
| diffusion-stable | pix2seq | wgan-gp-pytorch |
| diffwave-vocoder | pix2vox | whatsapp-gpt |
| discodiffusion | pix4d | world-models-a3c |
| disco-diffusion | pixabay-api | world-models-rl |
| disco-diffusion-local | pixel | world-vocoder |
| disco-diffusion-windows | pixel2style2pixel | yolo-model |
| discogan | pixelcnn | yolov3 |
| discrete-variational-autoencoders | pixel-cnn | yolov4 |
| docker-chatgpt | pixelrnn | yolov5 |
| doppelganger | pixelsnail | yolov6 |
| dragan | pixel-to-pixel-gan | yolov7 |
| dualgan | plugandplay | yolov8 |
| dvae | pokegan | |